%% file: main.tex
\documentclass[12pt]{article}
\usepackage{natbib,bm}
\usepackage{amsmath}
\usepackage{amssymb}
\usepackage{graphicx}
\usepackage[dvips]{epsfig}
\usepackage{lscape}

\renewcommand{\baselinestretch}{1.65}
\tiny\normalsize
\input Macros

\makeatletter
\def\vec{\mathop{\operator@font vec}\nolimits}
\makeatother

\DeclareMathOperator{\diag}{diag}
\DeclareMathOperator{\tr}{tr}

\begin{document}
\input text

\renewcommand{\baselinestretch}{1.1}

\tiny\normalsize
\bibliographystyle{agsm}
\bibliography{ref}

\newpage
\input table
\newpage
\input graph

\end{document}

%% file: Macros.tex
\newcommand{\nmathbf}{\bm}

\def\bfA{\nmathbf A}
\def\bfB{\nmathbf B}
\def\bfC{\nmathbf C}
\def\bfD{\nmathbf D}

\def\bfH{\nmathbf H}
\def\bfI{\nmathbf I}

\def\bfM{\nmathbf M}

\def\bfQ{\nmathbf Q}

\def\bfS{\nmathbf S}
\def\bfT{\nmathbf T}
\def\bfU{\nmathbf U}
\def\bfV{\nmathbf V}
\def\bfW{\nmathbf W}

\def\bfu{\nmathbf u}

\def\bfw{\nmathbf w}

\def\bfbeta   {\nmathbf \beta}
\def\bfgamma  {\nmathbf \gamma}

\def\bfDelta  {\nmathbf \Delta}

\def\bfLambda {\nmathbf \Lambda}

\def\bfSigma  {\nmathbf \Sigma}

\def\bfPhi    {\nmathbf \Phi}
\def\bfPsi    {\nmathbf \Psi}

\def\bfOmega  {\nmathbf \Omega}

\def\boldfacefake#1{\kern-4pt
   \hbox{ \mathsurround=0pt
   \hbox to 0.4pt{$#1$\hss}\hbox to 0.4pt{$#1$\hss}\hbox {$#1$}}}

\newcommand{\ok}{\hfill\fbox{}}

\newcommand{\btable}{\begin{table}[h]\centering}
\newcommand{\etable}{\end{table}}
\newcommand{\bt}{\begin{parag}\small \let\b=\nsb \let\sb=\nssb \begin{tabular}}
\newcommand{\et}{\end{tabular}\let\b=\nb \let\sb=\nsb\end{parag}}

\newenvironment{parag}{\par}{\par}
\newenvironment{dif}
    {\begin{parag}\small \let\b=\nsb \let\sb=\nssb \begin{parag}}
    {\let\b=\nb \let\sb=\nsb \end{parag}\end{parag}}

\newenvironment{proof}{\begin{dif} \noindent{\em Proof.~}}
            {\ok\vspace*{10pt}\end{dif}}
\newenvironment{exa}{\begin{list}{}
           {\setlength{\leftmargin}{10pt}
            \setlength{\rightmargin}{\leftmargin}}
           \item\begin{ex}\em}{\end{ex}\end{list}}

\newcommand{\be}{\begin{eqnarray}}
\newcommand{\ee}{\end{eqnarray}}
\newcommand{\ba}{\begin{eqnarray*}}
\newcommand{\ea}{\end{eqnarray*}}

\newtheorem{theorem0}{Theorem}
\newtheorem{lemma0}{Lemma}
\newtheorem{remark0}{Remark}
\newtheorem{fact0}{Fact}
\newtheorem{example0}{Example}
\newtheorem{definition0}{Definition}
\newtheorem{corollary0}{Corollary}
\newtheorem{proposition0}{Proposition}
\newtheorem{algorithmY}{Algorithm}

\newcommand{\reals}{\mbox{\rm I\kern-.20em R}}
\newcommand{\sreals}{\mbox{\small \rm I\kern-.20em R}}

\newcommand{\bdfn}{\begin{dfn}}
\newcommand{\edfn}{\end{dfn}}
\newcommand{\bteo}{\begin{teo}}
\newcommand{\eteo}{\end{teo}}
\newcommand{\bexa}{\begin{exa}}
\newcommand{\eexa}{\end{exa}}
\newcommand{\bdif}{\begin{dif}}
\newcommand{\edif}{\end{dif}}
\newcommand{\bpro}{\begin{proof}}
\newcommand{\epro}{\end{proof}}

%% file: text.tex
\begin{center}
 {\large \bf Graph-based Multivariate Conditional Autoregressive Models}

\bigskip
 Ye Liang \\
 {\small \it Department of Statistics, Oklahoma State University, Stillwater, Oklahoma 74078, U.S.A.} \\
 {\small ye.liang@okstate.edu}

\end{center}
\medskip

\begin{abstract}
The conditional autoregressive model is a routinely used statistical model for areal data that arise from, for instances, epidemiological, socio-economic or ecological studies. Various multivariate conditional autoregressive models have also been extensively studied in the literature and it has been shown that extending from the univariate case to the multivariate case is not trivial. The difficulties lie in many aspects, including validity, interpretability, flexibility and computational feasibility of the model. In this paper, we approach the multivariate modeling from an element-based perspective instead of the traditional vector-based perspective. We focus on the joint adjacency structure of elements and discuss graphical structures for both the spatial and non-spatial domains. We assume that the graph for the spatial domain is generally known and fixed while the graph for the non-spatial domain can be unknown and random. We propose a very general specification for the multivariate conditional modeling and then focus on three special cases, which are linked to well known models in the literature. Bayesian inference for parameter learning and graph learning is provided for the focused cases, and finally, an example with public health data is illustrated.

\end{abstract}

\newpage

\section{Introduction} \label{sec:intro}
Areal data, sometimes called lattice data, are usually represented by an undirected graph where each vertex
represents an areal unit and each edge represents a neighboring relationship. A finite set of random
variables on an undirected graph, where each vertex is a random variable, is called a Markov random 
field if it has the Markov property. Hence, the Markov random field models are often used for the areal data.
The univariate conditional autoregressive (\textsc{CAR}) model, originated from \cite{Besag:1974}, is 
a Gaussian Markov random field model, for which the joint distribution is multivariate Gaussian. Let $\bfu=(u_1,\ldots,u_I)^T$
be a vector of random variables on $I$ areal units (i.e. $I$ vertices). The zero-centered conditional autoregressive model
specifies full conditional Gaussian distributions
\ba
	u_i\mid u_{-i} \sim \mbox{N}\left( \sum_{i'\neq i} b_{ii'}u_{i'}, \tau_{i}^2 \right), ~~~ i=1,\ldots,I,
\ea
where $u_{-i}$ is the collection of $u_{i'}$ for $i'\neq i$. The resulting joint distribution, derived using Brook's lemma, 
has a density function as follows,
\ba
	f(\bfu\mid \bfT_{\textsc{CAR}},\bfB_{\textsc{CAR}})\propto \mbox{exp} \left\{ -\frac{1}{2}\bfu^T \bfT^{-1}_{\textsc{CAR}}
	(\bfI-\bfB_{\textsc{CAR}}) \bfu \right\},
\ea
where $\bfI$ is an identity matrix; $\bfB_{\textsc{CAR}}$ is an $I \times I$ matrix whose off-diagonal entries are $b_{ii'}$
and diagonal entries are zeros, and $\bfT_{\textsc{CAR}}=\diag\{\tau_{1}^2,\ldots,\tau_{I}^2\}$.
The joint distribution is multivariate Gaussian if and only if $\bfT^{-1}_{\textsc{CAR}} (\bfI-\bfB_{\textsc{CAR}})$ is symmetric and positive
definite. A further parameterization on $\bfB_{\textsc{CAR}}$ and $\bfT_{\textsc{CAR}}$ is needed to reduce the number of parameters in the model. Consider a so-called adjacency matrix $\bfC_{\textsc{CAR}}$ for the undirected graph, where the $ii'$th entry $C_{ii'}=1$ 
if unit $i$ and unit $i'$ are neighbors (denoted as $i\sim i'$) and $C_{ii'}=0$ otherwise.
One popular parameterization is to let $b_{ii'}=\rho C_{ii'}/C_{i+}$ and $\tau_i^2=\sigma^2/C_{i+}$, where
$C_{i+}$ is the $i$th row sum of $\bfC_{\textsc{CAR}}$, representing the total number of neighbors of unit $i$. 
Let $\bfD_{\textsc{CAR}}=\diag\{C_{1+},\ldots,C_{I+}\}$. When $\rho$ is strictly between the smallest and largest eigenvalues 
of $\bfD_{\textsc{CAR}}^{-1/2}\bfC_{\textsc{CAR}}\bfD_{\textsc{CAR}}^{-1/2}$, or sufficiently, when $|\rho|<1$, and $\sigma^2>0$, 
the joint distribution of $\bfu$ is a zero-mean multivariate Gaussian distribution:
$\bfu \sim N\{{\bf0}, \sigma^2(\bfD_{\textsc{CAR}}-\rho\bfC_{\textsc{CAR}})^{-1}\}$. This is called the proper conditional autoregressive model in the literature. When $\rho=1$, it is called the intrinsic conditional autoregressive model which is an improper distribution due to the singular covariance matrix. 

Turning to the multivariate case, consider $J$ responses (e.g. multiple diseases) on $I$ areal units. 
Let $\bfU$ be an $I \times J$ matrix-variate where the $ij$th entry $u_{ij}$ is a random variable for the  
$i$th areal unit and $j$th response. Each column of $\bfU$ is an areal vector for a single response and hence 
can be modeled by the univariate conditional autoregressive model. However, a multivariate model is desired for 
the matrix-variate $\bfU$ in order to simultaneously model the dependence across responses. 
Initially proposed by \cite{Mardia:1988}, the multivariate conditional autoregressive model specifies full 
conditional distributions on row vectors
of $\bfU$. Let $\bfu_i$ be the $i$th row vector of $\bfU$. Following \cite{Besag:1974}, specify
\be \label{mardia}
	\bfu_i\mid \bfu_{-i} \sim \mbox{N}\left(\sum_{i' \neq i} \bfB_{ii'}\bfu_{i'}, \bfSigma_i \right), ~~~ i=1,\ldots,I,
\ee
 where $\bfB_{ii'}$ and $\bfSigma_{i}$ are $J \times J$ matrices needing a further parameterization. 
 To make the joint distribution for $\vec(\bfU^T)$ a multivariate Gaussian, $\bfB_{ii'}$ and $\bfSigma_{i}$ must 
 satisfy certain conditions \citep{Mardia:1988}. \cite{Gelfand:2003} showed a convenient parameterization, 
 $\bfB_{ii'}=(\rho C_{ii'}/C_{i+})\bfI_J$ and $\bfSigma_i=\bfSigma/C_{i+}$. When $|\rho|<1$ and $\bfSigma$ is
 positive definite, $\vec(\bfU^T)$ has a zero-mean multivariate Gaussian distribution:
 $\vec(\bfU^T) \sim N\{{\bf0}, (\bfD_{\textsc{CAR}}-\rho\bfC_{\textsc{CAR}})^{-1}\otimes\bfSigma\}$. 
 It is clear that this multivariate specification is a Kronecker product formula where $(\bfD_{\textsc{CAR}}-\rho\bfC_{\textsc{CAR}})^{-1}$
 models the covariance structure across rows of $\bfU$ (spatial domain) and $\bfSigma$ models the covariance structure 
 across columns of $\bfU$ (response domain).
 From the modeling perspective, Mardia's specification has a difficulty with parameterization. 
 It is usually difficult to have a meaningful parameterization for $\bfB_{ii'}$ and $\bfSigma_i$ unless one 
 pursues a simple formulation. It is arguable that the Mardia's specification presents a conflict, where the between vector 
 variation is specified through an inverse covariance matrix, but the within vector variation is specified through 
 a covariance matrix. It seems more intuitive to either work with the joint covariance or the joint inverse covariance directly. 
 Notice that most multivariate spatial models for point reference data focus on the joint covariance structure. 
 In this paper, we focus on the joint inverse covariance structure of elements in the multivariate areal data. 
 In particular, we consider the joint adjacency structure of the lattice based on graphical structures of 
 both the spatial domain and the response domain. We build a framework for graph based multivariate conditional autoregressive models
 and discuss parameterizations under this framework. The advantage is that this framework is very general and we demonstrate
 it through multiple case examples. Furthermore, we allow graph learning for multiple responses in such models, which
 is potentially useful for many modern applications.   
 
 We shall point out other recent work on multivariate conditional autoregressive models. \cite{Kim:2001} and \cite{Jin:2005} 
 proposed conditional autoregressive models for bivariate areal data. Multivariate models were considered by 
 \cite{Gelfand:2003}, \cite{Jin:2007}, \cite{MacNab:2011, MacNab:2016}, \cite{Mart:2013, Mart:2017} among many others. 
 \cite{Macnab:2018} reviewed some recent developments on multivariate Gaussian Markov random field models. 
 We will show that some of the earlier work can be reconstructed in our proposed framework and some can be extended
 to graphical models. The paper is organized as follows. Section \ref{sec:spec} presents the general framework and three
 special parameterizations. Section \ref{sec:app} presents a real data example using the proposed models. 
 Section \ref{sec:dis} contains further discussions and remarks. Technical details are given in the appendix. 
 
\section{Graph-based Multivariate Conditional Autoregressive Models} \label{sec:spec}
\subsection{General framework}
Instead of specifying full conditional distributions on vectors like (\ref{mardia}), we approach this problem from 
an element-based perspective. Following \cite{Besag:1974}, specify full conditional distributions for each element $u_{ij}$ 
in the matrix-variate $\bfU$ as follows,
\ba
	u_{ij}\mid u_{-\{ij\}} \sim \mbox{N}\left( \sum_{\{i'j'\}\neq \{ij\}} b_{\{ij\},\{i'j'\}}u_{i'j'}, \tau_{ij}^2 \right),
	~~~ i=1,\ldots,I ~\mbox{and}~j=1,\ldots,J,
\ea
where $\{i'j'\}\neq \{ij\}$ means either $i'\neq i$ or $j'\neq j$. In fact, here we consider a lattice consisting of all elements in $\bfU$. 
Using Brook's lemma, the resulting joint distribution for $\vec(\bfU)$ is
\ba
	f(\vec(\bfU)\mid \bfB,\bfT)\propto \exp\left\{ -\frac{1}{2}\vec(\bfU)^T \bfT^{-1}(\bfI-\bfB)\vec(\bfU) \right\},
\ea
where $\bfI$ is an $IJ \times IJ$ identity matrix, 
$\bfT=\diag\{\tau_{11}^2,\ldots,\tau_{I1}^2,\ldots,\tau_{1J}^2,\ldots,\tau_{IJ}^2 \}$ and $\bfB$
can be expressed block-wisely,
\ba
	\bfB=\begin{pmatrix}
		\bfB_{11} & \cdots & \bfB_{1J} \\
		\vdots & \ddots & \vdots \\
		\bfB_{J1} & \cdots & \bfB_{JJ} 
	\end{pmatrix}
	\mbox{~~~where~~~}
	\bfB_{jj'}=
	\begin{pmatrix}
		b_{\{1j\},\{1j'\}} & \cdots & b_{\{1j\},\{Ij'\}} \\
		\vdots & \ddots & \vdots \\
		b_{\{Ij\},\{1j'\}} & \cdots & b_{\{Ij\},\{Ij'\}} 
	\end{pmatrix},	
\ea
and the diagonal elements $b_{\{ij\},\{ij\}}$ are zeros.
The joint distribution for $\vec(\bfU)$ is multivariate Gaussian if and only if $\bfT^{-1}(\bfI-\bfB)$ is symmetric
and positive definite. It is desired that $\bfB$ and $\bfT$ are further parameterized to reduce the number of parameters in the model. 
We denote this general model $\textsc{MCAR}(\bfB,\bfT)$ for later use.

Consider the adjacency structure of the undirected graph for the lattice of $\bfU$. In the univariate situation, the adjacency structure 
is determined by those geographical locations. Two areal units are connected by an edge if they are neighbors geographically. 
However, it is not obvious which elements should be neighbors in $\bfU$. Consider that the $J$ responses can be connected 
through an undirected graph. Let $\bfC^{(s)}$ be the adjacency matrix for all $I$ areal units and $\bfC^{(r)}$ be the adjacency matrix 
for all $J$ responses. 
Both the spatial graph and the response graph are then 
uniquely determined by $\bfC^{(s)}$ and $\bfC^{(r)}$, respectively. Let $\bfC$ be the joint adjacency matrix for
the lattice of $\bfU$. A general construction of $\bfC$ can be made through $\bfC^{(s)}$ and $\bfC^{(r)}$,
\be\label{adjacency}
	\bfC=\bfC^{(r)} \otimes \bfC^{(s)} + \bfC^{(r)} \otimes \bfI_I + \bfI_J \otimes \bfC^{(s)}.
\ee
This construction connects $u_{ij}$ with $u_{i'\sim i,j}$, $u_{i,j'\sim j}$ and $u_{i'\sim i, j'\sim j}$,
meaning its spatial neighbor, response neighbor and interaction neighbor, respectively. 
One may add edges for secondary neighbors or drop edges in a specific modeling. For example, some reduced constructions 
would be: (i) $\bfC=\bfI_J \otimes \bfC^{(s)}$ (independent conditional autoregressive models, 
no dependence between responses); (ii) $\bfC=\bfC^{(r)} \otimes \bfI_I$ (independent multivariate variables, 
no spatial dependence); (iii) $\bfC=\bfC^{(r)} \otimes \bfI_I + \bfI_J \otimes \bfC^{(s)}$ (drop edges for 
interaction neighbors $u_{i'\sim i, j'\sim j}$).  

Let $C_{\{ij\},\{i'j'\}}$ denote entries in $\bfC$, analogous to the block-wise notation $b_{\{ij\},\{i'j'\}}$ for $\bfB$.
Let  $d_j^{(r)}$ be the $j$th row sum in $\bfC^{(r)}$ and $d_i^{(s)}$ be the $i$th row sum in $\bfC^{(s)}$.
Then the $ij$th row sum in $\bfC$ is $d_{ij}=d_j^{(r)}d_i^{(s)}+d_j^{(r)}+d_i^{(s)}$.
Let $\bfD^{(r)}=\diag\{d_1^{(r)},\ldots,d_J^{(r)}\}$, $\bfD^{(s)}=\diag\{d_1^{(s)},\ldots,d_I^{(s)}\}$ and 
$\bfD=\diag\{d_{11},\ldots,d_{I1},\ldots,d_{1J},\ldots,d_{IJ}\}$. With the adjacency constructions and notations, 
we then explore further parameterization on $\bfB$ and $\bfT$ in the following subsections, and specifically, 
we discuss three specifications made from this general framework, all of which are linked to well known models in the literature.  

\subsection{Model 1: nonseparable multifold specification}
\cite{Kim:2001} developed a twofold conditional autoregressive model for bivariate areal data ($J=2$), 
using different linkage parameters for different types of neighbors. Those linkage parameters, in their work, are called smoothing
and bridging parameters, representing the strength of information sharing. If we extend their specification to 
an arbitrary $J$, we can parameterize $\bfB$ and $\bfT$ in the following way (assuming $i \neq i'$ and $j \neq j'$):
\ba
	&& b_{\{ij\},\{i'j\}} = \frac{\lambda_j}{d_{ij}} C_{\{ij\},\{i'j\}}, ~~~ 
		b_{\{ij\},\{ij'\}} = \frac{\psi_{jj'}}{d_{ij}} \sqrt{\frac{\delta_j}{\delta_{j'}}} C_{\{ij\},\{ij'\}},  \\
	&& b_{\{ij\},\{i'j'\}} = \frac{\phi_{jj'}}{d_{ij}} \sqrt{\frac{\delta_j}{\delta_{j'}}} C_{\{ij\},\{i'j'\}}, ~~~
		\tau_{ij}^2=\frac{\delta_j}{d_{ij}},
\ea
where $\lambda_j$, $\psi_{jj'}$ and $\phi_{jj'}$ are linkage parameters and $\delta_j$ are variance components.    
Linkage parameters are for three types of neighbor: $u_{i'\sim i,j}$, $u_{i,j'\sim j}$ and $u_{i'\sim i, j'\sim j}$.
Having this specification, the conditional mean of $u_{ij}$ essentially is
\ba
	\mbox{E}(u_{ij} \mid u_{-\{ij\}})=\frac{1}{d_{ij}}\left( \lambda_j\sum_{i' \sim i}u_{i'j}
	+\psi_{jj'}\sum_{j'\sim j}\sqrt{\frac{\delta_j}{\delta_{j'}}}u_{ij'}
	+\phi_{jj'}\sum_{i'\sim i}\sum_{j'\sim j}\sqrt{\frac{\delta_j}{\delta_{j'}}} u_{i'j'} \right),
\ea
which is a weighted average of all its neighbors in $\bfC$. This specification generalizes 
\cite{Kim:2001}' twofold model and hence could be called a multifold specification. 
It can be shown that, for this parameterization, the joint precision matrix is
\be \label{Model1}
	\bfT^{-1}(\bfI-\bfB) &=& (\bfDelta^{-\frac{1}{2}}\otimes\bfI_I ) \left\{\bfD-\bfLambda\otimes\bfC^{(s)} \right. \notag \\
	&& \left. -(\bfPsi\circ\bfC^{(r)})\otimes\bfI_I - (\bfPhi\circ\bfC^{(r)})\otimes\bfC^{(s)}\right\}
	(\bfDelta^{-\frac{1}{2}}\otimes\bfI_I ),
\ee
where $\bfDelta=\diag\{\delta_1,\ldots,\delta_J\}$, $\bfLambda=\diag\{\lambda_1,\ldots,\lambda_J\}$,
$\bfPsi$ and $\bfPhi$ are $J \times J$ symmetric matrices with entries $\psi_{jj'}$ and  $\phi_{jj'}$, respectively,
and the operator $\circ$ means an element-wise product. A derivation of (\ref{Model1}) is given in Appendix 1.
Note that only nonzero entries of $\bfPsi$ and $\bfPhi$ are parameters in the model, the number of which depends 
on $\bfC^{(r)}$.

In order to make (\ref{Model1}) positive definite, constraints on $\lambda_j$, $\psi_{jj'}$ and $\phi_{jj'}$ are needed,
assuming that $\delta_j>0$. In general, it is difficult to find a sufficient and necessary condition for the positive
definiteness of (\ref{Model1}). \cite{Kim:2001}'s solution to this problem 
was a sufficient condition: $\mbox{max}\{|\lambda_j|, |\psi_{jj'}|, |\phi_{jj'}|; \forall j, j'\}<1$, under which
the matrix (\ref{Model1}) is diagonally dominant and hence is positive definite. 
Though their proof was under $J=2$, it is true for any $J$ by the same arguments.
The advantage of this condition is that it is simple and implementable. However, this is not a
necessary condition meaning that it is impossible to reach all possible positive definite structures 
for the model under such a condition. 
In a Bayesian model, priors on parameters $\lambda_j$, $\psi_{jj'}$ and $\phi_{jj'}$ can be chosen based on
their actual constraints. In our case, a uniform prior $\mbox{Unif}(-1,1)$ is adequate for these linkage parameters. 
Priors on the variance components $\delta_j$ can be weakly-informative inverse-gamma priors $\mbox{IG}(a_j,b_j)$. 
Inference and computation under this model are given in Appendix 2.

\subsection{Model 2: separable specification with homogeneous spatial smoothing}
\cite{Gelfand:2003}'s Kronecker-product model is a convenient parameterization of \cite{Mardia:1988}'s
model. In our framework, this specification can be obtained and extended by having the following 
parameterization for $\bfB$ and $\bfT$ (assuming $i \neq i'$ and $j \neq j'$):
\ba
	&& b_{\{ij\},\{i'j\}} = \frac{\rho}{d_{i}^{(s)}} C_{\{ij\},\{i'j\}}, ~~~ 
		b_{\{ij\},\{ij'\}} = -\frac{\omega_{jj'}}{\omega_{jj}} C_{\{ij\},\{ij'\}}, \\
	&& b_{\{ij\},\{i'j'\}} = \frac{\rho\omega_{jj'}}{d_{i}^{(s)}\omega_{jj}} C_{\{ij\},\{i'j'\}}, ~~~
		\tau_{ij}^2=\frac{1}{d_{i}^{(s)}\omega_{jj}},
\ea
where $\rho$ and $\omega_{jj'}$ are linkage parameters, and $1/\omega_{jj}$ are variance components.
This parameterization does not seem straightforward, but is much clearer in the form
of conditional mean:
\be \label{cond_mean1}
	\mbox{E}\left( u_{ij}-\frac{\rho}{d_i^{(s)}}\sum_{i'\sim i}u_{i'j} ~\Bigg|~ u_{-\{ij\}} \right)
	=-\sum_{j'\sim j}\frac{\omega_{jj'}}{\omega_{jj}}\left( u_{ij'}-\frac{\rho}{d_i^{(s)}}\sum_{i'\sim i}u_{i'j'} \right).
\ee
Note that for a single response, the univariate conditional autoregressive model specifies 
$\mbox{E}(u_i | u_{-i})=\rho\sum_{i'\sim i} u_{i'}/d_i^{(s)}$. In the multivariate setting, 
$\rho\sum_{i'\sim i} u_{i'j}/d_i^{(s)}$ is no longer the conditional mean for $u_{ij}\mid u_{-\{ij\}}$
and their conditional difference is regressed on other differences through $\omega_{jj'}$. 
This parameterization yields the joint precision matrix
\be \label{Model2}
	\bfT^{-1}(\bfI-\bfB) &=& \left\{\bfOmega\circ(\bfI_J+\bfC^{(r)})\right\} \otimes (\bfD^{(s)}-\rho\bfC^{(s)}) \notag \\
	 &=&  \bfOmega_{\bfC^{(r)}} \otimes (\bfD^{(s)}-\rho\bfC^{(s)}),
\ee 
where $\bfOmega$ is a symmetric $J\times J$ matrix with entries $\omega_{jj'}$. A derivation of (\ref{Model2})
is given in Appendix 1. The linkage parameter
$\rho$ is interpreted as a spatial smoothing parameter and $\bfOmega$ controls the dependence across 
$J$ responses. It is noteworthy that only nonzero entries in $\bfOmega$ are parameters in the model and we denote
$\bfOmega_{\bfC^{(r)}}=\bfOmega\circ(\bfI_J+\bfC^{(r)})$ for simplicity. The notation $\bfOmega_{\bfC^{(r)}}$,
commonly used in graphical models, means the precision matrix restricted by graph $\bfC^{(r)}$. 
The model (\ref{Model2}) is a natural extension of \cite{Gelfand:2003}'s model.  When $\bfC^{(r)}$ is the complete graph 
(any two vertices are connected), $\bfOmega_{\bfC^{(r)}}$ is free of zero entries. Then let 
$\bfSigma=\bfOmega^{-1}$ and (\ref{Model2}) is equivalent to \cite{Gelfand:2003}'s specification.
We call this a completely separable specification because the Kronecker product completely separates
the spatial domain and the response domain. This complete separation is often not desirable because
it makes the spatial smoothing common for all $j$. We call this homogeneous spatial smoothing because
the linkage $\rho$ is the same for any $i$ and $i'$ which distinguishes Model 2 from Model 3 in the next subsection.
 
The joint precision matrix (\ref{Model2}) is positive definite if $|\rho|<1$ and $\bfOmega_{\bfC^{(r)}}$ is positive
definite. Let $M^+({\bfC^{(r)}})$ be the cone of symmetric positive definite matrices restricted by $\bfC^{(r)}$
and then $\bfOmega_{\bfC^{(r)}}\in M^+({\bfC^{(r)}})$. In a Bayesian model, a widely used prior on $\bfOmega_{\bfC^{(r)}}$
is the G-Wishart distribution \citep{Atay:2005,Letac:2007}. 
The G-Wishart distribution is a conjugate family for the precision matrix of a Gaussian graphical model, 
whose density function is given by
\ba
	p(\bfOmega_{\bfC^{(r)}} \mid b,\bfV)=I_{\bfC^{(r)}}(b,\bfV)^{-1}
	\left| \bfOmega_{\bfC^{(r)}} \right|^{\frac{b-2}{2}}\mbox{exp}
	\left\{-\frac{1}{2}\mbox{tr}(\bfV\bfOmega_{\bfC^{(r)}})\right\}
	1_{\bfOmega_{\bfC^{(r)}}\in M^+({\bfC^{(r)}})},
\ea
where $b>2$ is the number of degrees of freedom; $\bfV$ is the scale matrix and 
$I_{\bfC^{(r)}}(\cdot)$ is the normalizing constant. It is practically attractive
because of its conjugacy. That said, for a prior distribution $\mbox{GWis}(b,\bfV)$ and a given 
sample covariance matrix $\bfS$ of sample size $n$,
the posterior distribution of $\bfOmega_{\bfC^{(r)}}$ is $\mbox{GWis}(b+n,\bfV+\bfS)$.
Inference and computation under this model are given in Appendix 2.

\subsection{Model 3: separable specification with heterogenous spatial smoothing}
\cite{Dobra:2011} introduced a multivariate lattice model by giving Kronecker product G-Wishart priors
to the matrix-variate $\bfU$. In our framework, $\bfB$ and $\bfT$ can be parameterized in the following way
(assuming $i \neq i'$ and $j \neq j'$):
\ba
	&& b_{\{ij\},\{i'j\}} = -\frac{\omega_{ii'}^{(s)}}{\omega_{ii}^{(s)}} C_{\{ij\},\{i'j\}}, ~~~
		b_{\{ij\},\{ij'\}} = -\frac{\omega_{jj'}^{(r)}}{\omega_{jj}^{(r)}} C_{\{ij\},\{ij'\}}, \\
	&& b_{\{ij\},\{i'j'\}} = -\frac{\omega_{ii'}^{(s)}\omega_{jj'}^{(r)}}{\omega_{ii}^{(s)}\omega_{jj}^{(r)}} C_{\{ij\},\{i'j'\}}, ~~~ 
		\tau_{ij}^2=\frac{1}{\omega_{ii}^{(s)}\omega_{jj}^{(r)}},
\ea
which is equivalent to the version of conditional mean
\be \label{cond_mean2}
	\mbox{E}\left( u_{ij}-\frac{1}{\omega_{ii}^{(s)}}\sum_{i'\sim i}\omega_{ii'}^{(s)}u_{i'j} ~\Bigg|~ u_{-\{ij\}} \right)
	=-\sum_{j'\sim j}\frac{\omega_{jj'}}{\omega_{jj}}\left( u_{ij'}-\frac{1}{\omega_{ii}^{(s)}}\sum_{i'\sim i}
	\omega_{ii'}^{(s)}u_{i'j'} \right).
\ee
Comparing (\ref{cond_mean2}) with (\ref{cond_mean1}), instead of a homogeneous spatial smoothing 
with $\rho$, it has a heterogeneous specification with $\omega_{ii'}^{(s)}$. 
This is hence more flexible in the spatial domain. The resulting joint precision matrix is
\be \label{Model3}
	\bfT^{-1}(\bfI-\bfB) &=& \left\{\bfOmega^{(r)}\circ(\bfI_J+\bfC^{(r)})\right\} \otimes 
		\left\{\bfOmega^{(s)}\circ(\bfI_I+\bfC^{(s)})\right\} \notag \\
		&=& \bfOmega_{\bfC^{(r)}} \otimes \bfOmega_{\bfC^{(s)}} ,
\ee
where $\bfOmega^{(r)}$ is a symmetric $J \times J$ matrix with entries $\omega^{(r)}_{jj'}$ and 
$\bfOmega^{(s)}$ is a symmetric $I \times I$ matrix with entries $\omega^{(s)}_{ii'}$. A derivation of (\ref{Model3}) 
is given in Appendix 1. We again use $\bfOmega_{\bfC^{(r)}}$ and 
$\bfOmega_{\bfC^{(s)}}$ for simplicity. In model (\ref{Model2}), the spatial part is the conventional 
conditional autoregressive model while in model (\ref{Model3}), it is modeled by 
a more flexible one $\bfOmega_{\bfC^{(s)}}$. 

The precision matrix (\ref{Model3}) is positive definite if both $\bfOmega_{\bfC^{(r)}}$ and $\bfOmega_{\bfC^{(s)}}$
are positive definite. In a Bayesian model, both can have G-Wishart priors. The specification has an obvious 
problem of identification: $\bfOmega_{\bfC^{(r)}} \otimes  \bfOmega_{\bfC^{(s)}}=z\bfOmega_{\bfC^{(r)}} \otimes  
(1/z)\bfOmega_{\bfC^{(s)}}$, where $z$ is an arbitrary constant scalar. Following \cite{Wang:2009}, one can impose 
a constraint $\bfOmega_{\bfC^{(r)},11}=1$ and add an auxiliary variable $z$. Then specify a joint prior on
$(z,z\bfOmega_{\bfC^{(r)}})$:
\be \label{jointprior}
	p(z,z\bfOmega_{\bfC^{(r)}} \mid b^{(r)}, \bfV^{(r)})\propto p_{GWis}
	(z\bfOmega_{\bfC^{(r)}} \mid b^{(r)}, \bfV^{(r)})\cdot 1,
\ee
where $p_{GWis}(\cdot)$ is the density of G-Wishart distribution. Transform this joint density to $(z,\bfOmega_{\bfC^{(r)}})$
and we obtain the desired joint prior. There is no additional constraint imposed on $\bfOmega_{\bfC^{(s)}}$ and let 
$\bfOmega_{\bfC^{(s)}}\sim \mbox{GWis}(b^{(s)},\bfV^{(s)})$. Inference and computation under this model are given in Appendix 2.

\subsection{Priors for the graph}
The two types of graphs used in this modeling framework should be treated differently. On one hand, the spatial graph should be treated
known and fixed because the geographical locations and their neighboring structure is fixed in most scenarios. On the other hand,
the response graph should be treated unknown because we often know little about the relationship between multiple responses.  
In the literature of Gaussian graphical model determination, usually the unknown graph is assumed random and a prior on
the graph is assigned. The Markov chain Monte Carlo (MCMC) sampling scheme, such as the
reversible jump MCMC \citep{Green:1995}, is often used to sample graphs from the posterior distribution. 
In this paper, we adopt and slightly modify existing MCMC algorithms for the graph determination \citep{Wang:2012, Dobra:2011}, 
with computational details given in Appendix 2, for each aforementioned model. For the prior choice of $\bfC^{(r)}$, consider
\be \label{Gprior}
	P(\bfC^{(r)}) \propto B(a+\mbox{size}(\bfC^{(r)}), b+m-\mbox{size}(\bfC^{(r)}))/B(a,b),
\ee
where $B(\cdot,\cdot)$ is the beta function, $m$ is the total number of possible edges $J \choose 2$, 
$\mbox{size}(\bfC^{(r)}) \in \{0,1,\ldots,m\}$, and $a$ and $b$ are given hyperparameters. More details about this prior
can be found in \cite{Scott:2006} and \cite{Scott:2009}. The following prior is often used as well \citep{Dobra:2011}:
\be \label{Gprior2}
	P(\bfC^{(r)}) \propto \pi^{\mbox{size}(\bfC^{(r)})}(1-\pi)^{m-\mbox{size}(\bfC^{(r)})},
\ee
where $\pi \in (0,1)$ is a given hyperparameter. Sparser graphs can be favored by choosing a small value for $\pi$. 
The prior (\ref{Gprior}) can be obtained by integrating $\pi$ out with a hyperprior $\mbox{Beta}(a,b)$ on $\pi$. 

\section{An Application} \label{sec:app}
We illustrate the proposed models with a real example of disease mapping. It is known that smoking is linked with
multiple diseases in the population, of which leading diseases include lung diseases and heart diseases. 
The dataset under consideration here includes six variables, among which four variables are related to the smoke exposure 
and the other two are diseases. Obtained from the 2011 Missouri County Level Survey, the four smoke exposure variables are: 
Current Cigarette Smoking, Current Smokeless Tobacco Use, Current Other Tobacco Use, Exposure to Secondhand Smoke. 
Data are binary responses to the survey questionnaires (Yes or No), aggregated to each county level. 
The other two variables, obtained from the Surveillance, Epidemiology and End Results (SEER) program, are the 
Lung Cancer Mortality and the Heart Diseases Mortality, both of which are counts for each county within a specified time period.
To summarize, we have $I=115$ counties and $J=6$ response variables. Let $n_{i1},\ldots,n_{i4}$ be the numbers of respondents 
in the survey and let $E_{i5}$ and $E_{i6}$ be the age-adjusted expected mortality for the two diseases. 
Then, the proportions $y_{ij}/n_{ij},~j=1,\ldots,4$ are empirical estimates of the prevalences of the survey variables, 
and the proportions $y_{ij}/E_{ij},~j=5, 6$ are standardized mortality ratios of the diseases. 

Consider a Bayesian hierarchical model for $y_{ij}$. We use the binomial-logit model
and the Poisson-lognormal model \citep{Banerjee:2004} for $y_{i,1-4}$ and $y_{i,5-6}$, respectively, i.e.
\ba
	&& y_{ij} \sim \mbox{Bin}(n_{ij}, p_{ij}), ~~~ \mbox{logit}(p_{ij})=\beta_{j}+u_{ij}, ~~~ i=1,\ldots,115 \mbox{ and } j=1,\dots,4; \\
	&& y_{ij} \sim \mbox{Poi}(E_{ij}\eta_{ij}), ~~~ \mbox{log}(\eta_{ij})=\beta_{j}+u_{ij}, ~~~ i=1,\ldots,115 \mbox{ and } j=5, 6.
\ea
For simplicity, we do not consider other covariates in this example. The primary interest here is to model the random
effects $u_{ij}$, which are expected to be correlated in both the spatial domain and the response domain. 
To complete the model specification, specify a weakly-informative normal prior for the intercepts $\beta_j$ 
and a multivariate conditional autoregressive model $\textsc{MCAR}(\bfB,\bfT)$ for the random effects $\bfU=\{u_{ij}\}$. 
We apply the three proposed versions of $\textsc{MCAR}(\bfB,\bfT)$ here. Hyperparameters for prior distributions are
specified as follows. For the graph, noticing that the choice of $\pi$ in (\ref{Gprior2}) can influence the posterior inference, 
we consider the prior (\ref{Gprior2}) with both $\pi=0.2$ in favor of a sparse graph and $\pi=0.5$ as no preference. 
All other priors are chosen to be only weakly-informative and have little impact on the posterior inference. 
In Model 1, we specify hyperparameters in the inverse gamma prior as $a_j=b_j=0.5$. 
In Model 2, we specify hyperparameters in the G-Wishart prior as $b=3$ and $\bfV=\bfI$. 
In Model 3, we specify hyperparameters in the two G-Wishart priors as $b=3, \bfV=\bfI, b^{(s)}=24$ and 
$\bfV^{(s)}=(b^{(s)}-2)(\bfD-0.95\bfC)^{-1}$, which implies a prior mode at a proper conditional autoregressive model. 
For each model, we perform the Markov chain Monte Carlo for 150,000 iterations with a burn-in size of 50,000. 
Posterior results are based on the remaining samples. Figure \ref{fig:conv} shows the convergence of the log-joint-posterior 
and notice that they all converge quickly.  

Table \ref{tab:data} shows the posterior edge inclusion probabilities for the response graph $\bfC^{(r)}$. First, all three models seem to 
agree on the link between Cigarette Smoking and Secondhand Smoke Exposure, as well as the link between Lung Diseases Mortality
and Heart Diseases Mortality. There is a moderate agreement on the links between Secondhand Smoke Exposure and Lung Diseases
Mortality, and between Cigarette Smoking and Lung Diseases Mortality. In general, Model 1 tends to be a sparser graph, which 
is possibly due to the diagonal dominance condition. Model 2 is the simplest model as reflected by its {\it pD}, the effective number of 
parameters, but has the largest {\it DIC}. Model 3 is the most flexible model among the three, and the inferred graph tends to 
be denser than the other two. It is as expected that its {\it pD} is larger but the overall criterion {\it DIC} is much smaller than
the other two. Second, the edge inclusion probabilities are in general higher when $\pi=0.5$, as expected, but it has little material impact 
on the final inferred graph. The {\it DIC} has little change with different $\pi$ values. Lastly, Figures 1 - 3 show the maps of spatial random 
effects $u_{ij}$ for the three models, respectively, and for a problem of disease mapping, this is often the eventual output for practitioners. 

\section{Simulation} \label{sec:sim}
To validate the proposed algorithms, we perform a simulation study on a $7 \times 7$ regular grid ($I=49$ area units) with $J=4$ response variables. 
Consider the true response graph with two edges $C^{(r)}_{13}$ and $C^{(r)}_{24}$. In this simulation study, we do not consider the scenario 
with misspecified models, and therefore, data are generated under each of the three models and the correct model is then used for inference. 
The parameter settings are given as follows. For Model 1, $\lambda_j=0.95$, $\phi_{jj'}=\psi_{jj'}=0.9$, $\delta_j=1$ and $\beta_j=1$. 
For Model 2, $\rho=0.9$, $\omega_{jj}=4$, $\omega_{13}=\omega_{24}=-3.2$ and $\beta_j=1$. For Model 3, parameters are the same as Model 2 
but $\bfOmega_{\bfC^{(s)}}$ is generated from $\mbox{GWis}(10, 8(\bfD-0.9\bfC)^{-1})$. We repeat the simulation and inference process for $L=50$ times 
and for each time, the MCMC iteration number is 5,000. We consider three measures for validating and comparing the three algorithms. 
The first measure is the mean inclusion probability matrix with standard deviations. We call the second measure the error rate of mis-identified edges. 
If we use 0.5 as the threshold for identifying an edge in the graph, for each replication, we obtain an inferred graph and then compare with the true graph 
to record a proportion of wrong edges/non-edges. The error rate is the average proportion of $L$ replications. The third measure is the mean absolute 
error (MAE) of random effects in the model, 
\ba
	\mbox{MAE}=\frac{1}{L}\frac{1}{J}\frac{1}{I}\sum_l \sum_j \sum_i \left|\frac{\hat{u}_{ijl}-u_{ijl}}{u_{ijl}}\right|
\ea
where $u_{ijl}$ is the true value and $\hat{u}_{ijl}$ is the posterior mean. 

Simulation results are given in Table \ref{tab:sim}. For all three models, the algorithms can correctly identify the true edges. The algorithm for 
Model 1 appears to be unstable as the standard deviation is large and tends to underestimate inclusion probabilities, while the algorithm for Model 3 
tends to overestimate inclusion probabilities for non-edges. The algorithm for Model 2 presents the smallest error rate and MAE. Note that this simulation 
study validates the proposed algorithms under correct model specifications and hence the result cannot imply that Model 2 is the best model 
for a real dataset. In fact, as shown in the data analysis, Model 2 is the simplest specification and is the least preferred model in that case according to {\it DIC}. 

\section{Further Discussion} \label{sec:dis}
In this paper, we proposed a modeling framework for multivariate areal data from a graphical model perspective. 
We rebuilt three well known models in our framework and developed Bayesian inference tools for the proposed models. 
It is our perspective that this framework is very general and can contain other models that are beyond the cases 
discussed in the paper. For example, \cite{Jin:2007} specified a co-regionalized areal data model, in which their Case 3 is 
a very general specification. We show that this specification can be reproduced and extended in our framework. 
Consider the Cholesky decomposition $\bfSigma=\bfA\bfA^T$. \cite{Jin:2007}'s Case 3 specification of the joint covariance matrix is
$(\bfA\otimes\bfI_I)(\bfI_J\otimes\bfD^{(s)}-\bfPhi\otimes\bfC^{(s)})^{-1}(\bfA\otimes\bfI_I)^T$ whose inverse is then
\be \label{Jin}
	(\bfA\bfA^T)^{-1}\otimes\bfD^{(s)}-(\bfA^{-1})^{T}\bfPhi\bfA^{-1} \otimes \bfC^{(s)}
\ee   
where $\bfPhi$ is a symmetric $J\times J$ matrix. Let $\bfOmega=\bfSigma^{-1}=(\bfA\bfA^T)^{-1}$ and
$\bfQ=(\bfA^{-1})^{T}\bfPhi\bfA^{-1}$. Obviously it is one-to-one from $(\bfA,\bfPhi)$ to $(\bfOmega,\bfQ)$.
Specification (\ref{Jin}) is hence equivalent to 
\be \label{Jin_prec}
	\bfOmega\otimes\bfD^{(s)}-\bfQ\otimes\bfC^{(s)},
\ee 
where $\bfQ$ is a symmetric $J\times J$ matrix with entries $q_{jj'}$. 
To reproduce this specification in our framework, parameterize $\bfB$ and $\bfT$ as follows 
(assuming $i \neq i'$ and $j \neq j'$):
\ba
	&& b_{\{ij\},\{i'j\}} = \frac{q_{jj}}{d_{i}^{(s)}\omega_{jj}} C_{\{ij\},\{i'j\}}, ~~~ 
		b_{\{ij\},\{ij'\}} = -\frac{\omega_{jj'}}{\omega_{jj}} C_{\{ij\},\{ij'\}}, \\
	&& b_{\{ij\},\{i'j'\}} = \frac{q_{jj'}}{d_{i}^{(s)}\omega_{jj}} C_{\{ij\},\{i'j'\}}, ~~~
		\tau_{ij}^2=\frac{1}{d_{i}^{(s)}\omega_{jj}}.
\ea
This parameterization leads to the joint precision matrix
\be \label{Model4}
	\bfT^{-1}(\bfI-\bfB) &=& \left\{\bfOmega\circ(\bfI_J+\bfC^{(r)})\right\} \otimes \bfD^{(s)}-\left\{\bfQ\circ(\bfI_J+\bfC^{(r)})\right\} \otimes\bfC^{(s)} \notag \\
	&=&  \bfOmega_{\bfC^{(r)}} \otimes \bfD^{(s)}-\bfQ_{\bfC^{(r)}} \otimes\bfC^{(s)}.
\ee
The expression (\ref{Model4}) reduces to (\ref{Jin_prec}) which is equivalent to \cite{Jin:2007}'s (\ref{Jin}) 
when $\bfC^{(r)}$ is a complete graph. A derivation of (\ref{Model4}) is given in Appendix 1. The validity of this model 
relies on the positive definiteness of (\ref{Model4}). \cite{Jin:2007} showed that it is positive
definite if $\bfOmega$ is positive definite and eigenvalues of $\bfPhi=\bfA^T\bfQ\bfA$
are between $1/\xi_{\mbox{min}}$ and $1/\xi_{\mbox{max}}$, reciprocals of the smallest and largest eigenvalues of 
$\bfD^{(s)-1/2}\bfC^{(s)}\bfD^{(s)-1/2}$, which are known constants. The graphical version (\ref{Model4}) must also
satisfy this condition, that is, $1/\xi_{\mbox{min}} \leq \lambda(\bfOmega_{\bfC^{(r)}}^{-1}\bfQ_{\bfC^{(r)}})\leq 1/\xi_{\mbox{max}}$, 
where $\lambda(\bfM)$ is any eigenvalue of $\bfM$. Considering that both $\bfOmega_{\bfC^{(r)}}$ and $\bfQ_{\bfC^{(r)}}$
are restricted by the underlying graph, the eigenvalue condition is not easy to implement in computations. 
This matter is worth investigating in the future. 

In general, flexible models are desired for modeling multivariate areal data because overly simplistic models may misspecify the
true underlying covariance structure. However, there is almost always a trade-off between
the simplicity and the flexibility of a model. It is probably reasonable to allow certain flexibilities for specific purposes, such as 
in this paper, for learning a graphical relationship between multiple responses. It is usually the practitioner's choice whether
a more flexible but complicated model is needed for the problem at hand, especially when the performance improvement is negligible. 

\section*{Appendix 1: Derivations} \label{app:a1}
\subsection*{Derivation of equation (\ref{Model1})} 
With the parameterization in Model 1, we have $\bfT=\bfD^{-1}(\bfDelta \otimes \bfI_I)$ and 
\ba
	\bfB=(\bfDelta^{\frac{1}{2}}\otimes\bfI_I)\bfD^{-1}\left\{\bfLambda\otimes\bfC^{(s)}+(\bfPsi\circ\bfC^{(r)})\otimes\bfI_I 
		+ (\bfPhi\circ\bfC^{(r)})\otimes\bfC^{(s)}\right\}(\bfDelta^{-\frac{1}{2}}\otimes\bfI_I).
\ea
Then immediately we have expression (\ref{Model1}) for $\bfT^{-1}(\bfI-\bfB)$. 

\subsection*{Derivation of equation (\ref{Model2})}
With the parameterization in Model 2, we have $\bfT=\left(\diag(\bfOmega) \otimes \bfD^{(s)}\right)^{-1}$ and
\ba
	\bfB=\rho\bfI_J\otimes\bfD^{(s)-1}\bfC^{(s)}-\left(\diag(\bfOmega)^{-1}\bfOmega\circ\bfC^{(r)}\right)\otimes\bfI_I
		+\rho\left(\diag(\bfOmega)^{-1}\bfOmega\circ\bfC^{(r)}\right)\otimes\bfD^{(s)-1}\bfC^{(s)}.
\ea
Then
\ba
	\bfT^{-1}(\bfI-\bfB) &=& \diag(\bfOmega)\otimes \bfD^{(s)}-\rho\diag(\bfOmega)\otimes\bfC^{(s)}
		+(\bfOmega\circ\bfC^{(r)})\otimes\bfD^{(s)}-\rho(\bfOmega\circ\bfC^{(r)})\otimes\bfC^{(s)} \\
		&=& \left\{\bfOmega\circ(\bfI_J+\bfC^{(r)})\right\}\otimes\bfD^{(s)}- \left\{\bfOmega\circ(\bfI_J+\bfC^{(r)})\right\}\otimes\rho\bfC^{(s)} \\
		&=&  \left\{\bfOmega\circ(\bfI_J+\bfC^{(r)})\right\}\otimes(\bfD^{(s)}-\rho\bfC^{(s)}),
\ea
which is expression (\ref{Model2}).

\subsection*{Derivation of equation (\ref{Model3})}
With the parameterization in Model 3, we have $\bfT=\left(\diag(\bfOmega^{(r)})\otimes\diag(\bfOmega^{(s)})\right)^{-1}$ and
\ba
	\bfB &=& -\bfI_J\otimes\left(\diag(\bfOmega^{(s)})^{-1}\bfOmega^{(s)}\circ\bfC^{(s)}\right)-
		\left(\diag(\bfOmega^{(r)})^{-1}\bfOmega^{(r)}\circ\bfC^{(r)}\right)\otimes\bfI_I \\
		&& -\left(\diag(\bfOmega^{(r)})^{-1}\bfOmega^{(r)}\circ\bfC^{(r)}\right)\otimes
		\left(\diag(\bfOmega^{(s)})^{-1}\bfOmega^{(s)}\circ\bfC^{(s)}\right).
\ea
Then
\ba
	\bfT^{-1}(\bfI-\bfB) &=& \diag(\bfOmega^{(r)})\otimes\diag(\bfOmega^{(s)})
		+ \diag(\bfOmega^{(r)})\otimes(\bfOmega^{(s)}\circ\bfC^{(s)}) \\
		&& + (\bfOmega^{(r)}\circ\bfC^{(r)})\otimes\diag(\bfOmega^{(s)})
			+ (\bfOmega^{(r)}\circ\bfC^{(r)})\otimes(\bfOmega^{(s)}\circ\bfC^{(s)}) \\
		&=& \left\{\bfOmega^{(r)}\circ(\bfI_J+\bfC^{(r)})\right\} \otimes 
			\left\{\bfOmega^{(s)}\circ(\bfI_I+\bfC^{(s)})\right\},
\ea
which is expression (\ref{Model3}).

\subsection*{Derivation of equation (\ref{Model4})}
With the parameterization in Model 4, we have $\bfT=\left(\diag(\bfOmega) \otimes \bfD^{(s)}\right)^{-1}$ and
\ba
	\bfB &=& \diag(\bfOmega)^{-1}\diag(\bfQ) \otimes \bfD^{(s)-1}\bfC^{(s)}
	-\diag(\bfOmega)^{-1}\bfOmega\circ \bfC^{(r)} \otimes \bfI_I \\
	&& +\diag(\bfOmega)^{-1}\bfQ\circ \bfC^{(r)} \otimes \bfD^{(s)-1}\bfC^{(s)}.
\ea
Then
\ba
	\bfT^{-1}(\bfI-\bfB) &=& \diag(\bfOmega)\otimes\bfD^{(s)} - \diag(\bfQ)\otimes\bfC^{(s)}
		+\bfOmega\circ\bfC^{(r)}\otimes\bfD^{(s)} - \bfQ\circ\bfC^{(r)}\otimes\bfC^{(s)} \\
		&=& \left\{\bfOmega\circ(\bfI_J+\bfC^{(r)})\right\} \otimes \bfD^{(s)}-\left\{\bfQ\circ(\bfI_J+\bfC^{(r)})\right\} \otimes\bfC^{(s)},
\ea
which is expression (\ref{Model4}).

\section*{Appendix 2: Bayesian Computations} \label{app:a2}
\subsection*{A hierarchical generalized linear model}
For illustration, we now assume a full Bayesian hierarchical model and give computational details 
for this model. Assume binomial counts $y_{ij}/n_{ij}$ for $J$ responses and $I$ areal units. 
Specify a Bayesian model as follows, for $i=1,\ldots,I$ and $j=1,\dots,J$,
\ba
	&& y_{ij} \mid p_{ij} \sim \mbox{Bin}(n_{ij}, p_{ij}), ~~~~~~~~ \mbox{logit}(p_{ij})=\beta_j+u_{ij}, \\
	&& \beta_j \sim \mbox{N}(0, \tau_0^2), ~~~~~~~~ \bfU \mid \bfB,\bfT \sim \textsc{MCAR}(\bfB,\bfT),
\ea
where $\tau_0^2$ is a given constant, $\bfU$ is the matrix-variate of $u_{ij}$,
and priors for $\bfB$ and $\bfT$ depend on the specific parameterization.
This section is organized as follows: we first give details of updating effects parameters $\beta_j$ and $u_{ij}$,
and then, separately for each model, details of updating parameters of $\textsc{MCAR}$ and updating the random response
graph $\bfC^{(r)}$.

\subsection*{Updating effects parameters}
Our experience has shown that the convergence is poor if we directly update $\beta_j$ and $u_{ij}$. 
We apply the hierarchical centering technique \citep{Gelfand:1995} and block sampling.
Let $\gamma_{ij}=\beta_j+u_{ij}$ and $\bfgamma=\vec[(\gamma_{ij})_{I\times J}]$ has a non-centered 
\textsc{MCAR} prior. We update $(\gamma_{ij},\beta_j)$ instead of $(u_{ij},\beta_j)$. The full conditional distribution of $\gamma_{ij}$ is
\ba
	p(\gamma_{ij}\mid\cdot)\propto \frac{e^{y_{ij}\gamma_{ij}}}{(1+e^{\gamma_{ij}})^{n_{ij}}}
	\exp \left\{ -\frac{1}{2\tau_{ij}^2}\left( \gamma_{ij}-\beta_j-\sum_{\{i'j'\}\neq \{ij\}} b_{\{ij\},\{i'j'\}}(\gamma_{i'j'}-\beta_{j'}) \right) \right\}.
\ea
We use Metropolis-Hastings algorithm to sample $\gamma_{ij}$ from this conditional density. 
We block sample $\bfbeta$ in the following way. For now denote $\bfM=\bfT^{-1}(\bfI-\bfB)$, the joint precision matrix.
Let $\bfgamma^*=\bfM\bfgamma$ and $\bfgamma^{**}$ be a $J \times 1$ vector such that $\gamma^{**}_1$
is the sum of the first $I$ elements in $\bfgamma^*$, $\gamma^{**}_2$ is the sum of the second $I$ elements 
in $\bfgamma^*$ and so on. Partition $\bfM$ into $J \times J$ blocks and define
\ba
	\bfH=\begin{pmatrix}
		{\bf1}^T\bfM_{11}{\bf1} & \cdots & {\bf1}^T\bfM_{1J}{\bf1} \\
		\vdots & \ddots & \vdots \\
		{\bf1}^T\bfM_{J1}{\bf1} & \cdots & {\bf1}^T\bfM_{JJ}{\bf1},
	\end{pmatrix}
\ea
where ${\bf1}$ is the all-one vector. Then the full conditional distribution for the vector $\bfbeta$ is
\ba
	(\bfbeta \mid \cdot) \sim \mbox{N}\left[ (\bfH+1/\tau_0^2 \bfI)^{-1}\bfgamma^{**}, (\bfH+1/\tau_0^2 \bfI)^{-1} \right].
\ea

\subsection*{Model 1: updating $\delta_j$, $\lambda_j$, $\psi_{jj'}$, $\phi_{jj'}$ and $\bfC^{(r)}$}
Given the current graph $\bfC^{(r)}$, parameters are updated through Gibbs sampling. 
Recall priors on these parameters: $\delta_j\sim \mbox{IG}(a_j,b_j)$ and $\lambda_j, \psi_{jj'}, \phi_{jj'}\sim \mbox{Unif}(-1,1)$.
Let $\bfu_j$ be the $j$th column vector of $\bfU$, $j=1,\ldots,J$ and $\bfD_j$ be the $j$th diagonal block of $\bfD$.
The full conditional distribution of $\delta_j$ is given by
\ba
	p(\delta_j\mid \cdot)\propto \delta_j^{-\frac{I}{2}-a_j-1} \exp\left\{ -\frac{1}{2\delta_j}\bfu_j^T
	(\bfD_j-\lambda_j\bfC^{(s)})\bfu_j + \sum_{j' \sim j} \frac{1}{\sqrt{\delta_j\delta_{j'}}}
	\bfu_j^T (\psi_{jj'}\bfI_I+\phi_{jj'}\bfC^{(s)}) \bfu_{j'} - \frac{b_j}{\delta_j} \right\}.
\ea
It can be shown that the transformed one $(\sqrt{1/\delta_j}\mid \cdot)$ is log-concave when $I+2a_j-1>0$. 
Thus, we use the adaptive rejection sampling to update $\delta_j$.

Let $\bfW$ be an $I \times J$ matrix, where $\vec(\bfW)=(\bfDelta^{-1/2}\otimes \bfI_I)\vec(\bfU)$ and 
$\bfw_j$ be the $j$th column vector of $\bfW$. Let $\bfM=\bfD-\bfLambda\otimes\bfC^{(s)}-(\bfPsi\circ\bfC^{(r)})\otimes\bfI_I 
- (\bfPhi\circ\bfC^{(r)})\otimes\bfC^{(s)}$ as in (\ref{Model1}). Then $\lambda_j$, $\psi_{jj'}$ and $\phi_{jj'}$ are 
sequentially updated through following full conditional distributions,
\ba
	p(\lambda_j \mid \cdot) &\propto& \left|\bfM(\lambda_j) \right|^{\frac{1}{2}}
		\exp\left( \frac{1}{2}\lambda_j\bfw_j^T \bfC^{(s)}\bfw_j \right), \\
	p(\psi_{jj'} \mid \cdot) &\propto& \left|\bfM(\psi_{jj'}) \right|^{\frac{1}{2}}
		\exp\left( \psi_{jj'}\bfw_j^T \bfw_{j'} \right), \\
	p(\phi_{jj'} \mid \cdot) &\propto& \left|\bfM(\phi_{jj'}) \right|^{\frac{1}{2}}
		\exp\left( \phi_{jj'}\bfw_j^T \bfC^{(s)} \bfw_{j'} \right).
\ea
We use Metropolis-Hastings algorithm to update these parameters. Note that evaluating the sparse $|\bfM|$ could be 
computationally intensive. An efficient algorithm, usually based on the Cholesky decomposition, on sparse matrices is 
helpful. 

The graph $\bfC^{(r)}$ is updated through a simple reversible jump \textsc{MCMC} algorithm. Propose a new graph $\bfC^{(r)*}$ by
only adding or deleting one edge from $\bfC^{(r)}$. Without loss of generality, suppose that one edge $\{j0, k0\}$ is added to the new graph.
Dimension has been changed by $2$ from $(\bfC^{(r)},\bfPsi,\bfPhi)$ to $(\bfC^{(r)*},\bfPsi^*,\bfPhi^*)$. Propose $u_1\sim U(-1,1)$
and $u_2\sim U(-1,1)$, and let $\psi^*_{j0,k0}=u_1$ and $\phi^*_{j0,k0}=u_2$. The Jacobian from $(\bfPsi,\bfPhi,u_1,u_2)$ to
$(\bfPsi^*, \bfPhi^*)$ hence is $1$. Choose a Bernoulli jump proposal with odds 
$q(\bfC^{(r)*}, \bfC^{(r)})/q(\bfC^{(r)}, \bfC^{(r)*})=p(\bfC^{(r)})/p(\bfC^{(r)*})$ and systematically scan through the graph for updating. 
Accept the move from $\bfC^{(r)}$ to $\bfC^{(r)*}$ with probability $\min\{1, \alpha\}$ where 
\ba
	\alpha=\frac{|\bfM^*|^{\frac{1}{2}}}{|\bfM|^{\frac{1}{2}}} \exp\left\{ -\frac{1}{2}\vec(\bfW)(\bfM^*-\bfM)\vec(\bfW) \right\}.
\ea

\subsection*{Model 2: updating $\rho$, $\bfOmega_{\bfC^{(r)}}$ and $\bfC^{(r)}$}
Given the current graph $\bfC^{(r)}$, parameters are updated through Gibbs sampling.
Recall priors on these parameters: $\rho\sim U(-1,1)$ and $\bfOmega_{\bfC^{(r)}}\sim \mbox{GWis}(b,\bfV)$.
Use Metropolis-Hastings algorithm to update $\rho$. It can be shown that the full conditional distribution for $\rho$ is
\ba
	p(\rho \mid \cdot) \propto \left| \bfD^{(s)}-\rho\bfC^{(s)} \right|^{\frac{J}{2}}
	\exp\left\{ \frac{\rho}{2}\vec(\bfU)^T (\bfOmega_{\bfC^{(r)}} \otimes \bfC^{(s)}) \vec(\bfU) \right\}.
\ea

Let $\bfW_1$ be an $I \times J$ matrix, where $\vec(\bfW_1)=[\bfI_J \otimes (\bfD^{(s)}-\rho\bfC^{(s)})^{1/2}]\vec(\bfU)$ and 
$\bfw_{1,j}$ be the $j$th column vector of $\bfW_1$. Let $\bfS$ be an $J\times J$ matrix with $s_{jj'}=\bfw_{1,j}^T \bfw_{1,j'}$. 
Then the full conditional distribution of $\bfOmega_{\bfC^{(r)}}$ is 
\ba
	p(\bfOmega_{\bfC^{(r)}} \mid \cdot)\propto \left| \bfOmega_{\bfC^{(r)}} \right|^{\frac{b+I-2}{2}} 
	\exp\left[ -\frac{1}{2} \tr\left\{ \bfOmega_{\bfC^{(r)}}(\bfV+\bfS) \right\} \right]
	1_{\bfOmega_{\bfC^{(r)}}\in M^+({\bfC^{(r)}})},
\ea
which is GWis($b+I, \bfV+\bfS$). For sampling from the G-Wishart distribution, we use the block Gibbs sampler, 
given the set of maximum cliques, introduced by \cite{Wang:2012}.

The graph $\bfC^{(r)}$ is updated using \cite{Wang:2012}'s partial analytic structure algorithm (p. 188, Algorithm 2).  

\subsection*{Model 3: updating $\bfOmega_{\bfC^{(r)}}$, $\bfOmega_{\bfC^{(s)}}$ and $\bfC^{(r)}$}
Given the current graph $\bfC^{(r)}$, parameters are updated through Gibbs sampling.
Recall that we impose a constraint and use a joint prior (\ref{jointprior}) on $(z,z\bfOmega_{\bfC^{(r)}})$ and have 
$\bfOmega_{\bfC^{(s)}}\sim \mbox{GWis}(b^{(s)},\bfV^{(s)})$. Let both $\bfW^{(r)}$ and $\bfW^{(s)}$ be $I \times J$ matrices,
where $\vec(\bfW^{(r)})=(\bfI_J \otimes \bfOmega_{\bfC^{(s)}}^{1/2})\vec(\bfU)$ and 
$\vec(\bfW^{(s)})=(\bfOmega_{\bfC^{(r)}}^{1/2} \otimes \bfI_I)\vec(\bfU)$. Let $\bfw_j^{(r)}$ be the $j$th column vector of $\bfW^{(r)}$
and $\bfw_i^{(s)}$ be the $i$th row vector of $\bfW^{(s)}$. Then let $\bfS^{(r)}$ be $J\times J$ with $s_{jj'}^{(r)}=\bfw_j^{(r)T}\bfw_j^{(r)}$
and $\bfS^{(s)}$ be $I\times I$ with $s_{ii'}^{(s)}=\bfw_i^{(s)T}\bfw_i^{(s)}$. With these notations, we have
\ba
	(z\mid \cdot) \sim Ga(a_z,b_z),
\ea 
where $a_z=J(b-2)/2+\nu(\bfC^{(r)})$ and $b_z=\tr(\bfOmega_{\bfC^{(r)}}\bfV^{(r)})/2$;
\ba
	(\bfOmega_{\bfC^{(r)}}\mid \cdot) \sim \mbox{GWis}(b^{(r)}+I, z\bfV^{(r)}+\bfS^{(r)})
\ea
and 
\ba
	(\bfOmega_{\bfC^{(s)}}\mid \cdot) \sim \mbox{GWis}(b^{(s)}+J, \bfV^{(s)}+\bfS^{(s)}).
\ea

The graph $\bfC^{(r)}$ is updated using \cite{Wang:2012}'s partial analytic structure algorithm (p. 188, Algorithm 2).

%% file: table.tex
\begin{table} 
\begin{center}
\begin{tabular}{c|cccccc|cc}
\hline
\hline
\multicolumn{8}{c}{Model 1} \\
\hline
 & Cigarette & Smokeless & Other & Secondhand & Lung & Heart & pD & DIC \\
\hline 
Cigarette & & 0 & 0.165 & {\bf 0.963} & 0.276 & 0 & 460.8 & 5766.5  \\
Smokeless & 0 & & 0.181 & 0 & 0 & 0.002 & & \\
Other & 0.258 & 0 & & 0.032 & 0.014 & 0.005 & & \\
Secondhand & {\bf 0.983} & 0 & 0.062 & & 0.453 & 0 & & \\
Lung & 0.301 & 0 & 0.040 & {\bf 0.544} & & {\bf 0.533} & & \\
Heart & 0.005 & 0 & 0.039 & 0.007 & {\bf 0.772} & & 460.2 & 5765.0 \\
\hline
\hline
\multicolumn{8}{c}{Model 2} \\
\hline
 & Cigarette & Smokeless & Other & Secondhand & Lung & Heart & pD & DIC \\
\hline 
Cigarette & & 0.213 & {\bf 0.919} & {\bf 1} & {\bf 0.756} & 0.373 & 445.1 & 5813.8 \\
Smokeless & 0.340 & & 0.207 & 0.281 & 0.249 & {\bf 0.706} & & \\
Other & {\bf 0.872} & 0.316 & & 0.390 & 0.346 & {\bf 0.898} & & \\
Secondhand & {\bf 1} & {\bf 0.527} & 0.475 & & {\bf 0.732} & 0.343 & & \\
Lung & {\bf 0.584} & 0.407 & 0.412 & {\bf 0.871} & & {\bf 1} & & \\
Heart & 0.352 & {\bf 0.942} & {\bf 0.838} & 0.400 & {\bf 1} & & 447.2 & 5812.6 \\
\hline
\hline
\multicolumn{8}{c}{Model 3} \\
\hline
 & Cigarette & Smokeless & Other & Secondhand & Lung & Heart & pD & DIC \\
\hline 
Cigarette & & 0.268 & 0.328 & {\bf 0.827} & {\bf 0.541} & 0.436 & 543.3 & 5630.4 \\
Smokeless & 0.249 & & 0.417 & {\bf 0.692} & {\bf 0.515} & {\bf 0.726} & & \\
Other & 0.317 & 0.403 & & {\bf 0.534} & 0.453 & {\bf 0.510} & & \\
Secondhand & {\bf 0.819} & {\bf 0.694} & {\bf 0.531} & & {\bf 0.936} & {\bf 0.801} & & \\
Lung & {\bf 0.521} & {\bf 0.500} & 0.435 & {\bf 0.941} & & {\bf 0.997} & & \\
Heart & 0.411 & {\bf 0.732} & {\bf 0.508} & {\bf 0.805} & {\bf 0.998} & & 543.1 & 5629.2 \\
\hline
\end{tabular}
\end{center}
\caption{Data analysis: Posterior edge inclusion probabilities for the response graph and deviance information criterion for Models 1 - 3 respectively. Inclusion probabilities higher than 0.5 are in bold. The upper-right panel is for $\pi=0.2$ and the lower-left panel is for $\pi=0.5$. }
\label{tab:data}
\end{table}

\begin{table} 
\begin{center}
\begin{tabular}{c|ccc|cc}
\hline
\hline
\multicolumn{5}{c}{Model 1} \\
\hline
 & Var 2 & Var 3 & Var 4 & Error Rate & MAE \\
\hline 
 Var 1 & 0.128 (0.214) & {\bf 0.664} (0.358) & 0.088 (0.107) & 0.15 & 3.167 \\
 Var 2 & & 0.100 (0.148) & {\bf 0.615} (0.352) & & \\
 Var 3 & & & 0.087 (0.155) & & \\
\hline
\hline
\multicolumn{5}{c}{Model 2} \\
\hline
 & Var 2 & Var 3 & Var 4 & Error Rate & MAE \\
\hline 
 Var 1 & 0.211 (0.108) & {\bf 1} (0) & 0.200 (0.086) & 0.02 & 2.153 \\
 Var 2 & & 0.213 (0.134) & {\bf 1} (0) & & \\
 Var 3 & & & 0.199 (0.107) & & \\
\hline
\hline
\multicolumn{5}{c}{Model 3} \\
\hline
 & Var 2 & Var 3 & Var 4 & Error Rate & MAE \\
\hline 
 Var 1 & 0.246 (0.037) & {\bf 0.961} (0.051) & 0.292 (0.054) & 0.07 & 2.218 \\
 Var 2 & & 0.438 (0.092) & {\bf 0.996} (0.008) & & \\
 Var 3 & & & 0.426 (0.083) & & \\
\hline
\end{tabular}
\end{center}
\caption{Simulation: Mean posterior edge inclusion probabilities (standard deviations in parentheses), error rates of mis-identified edges, and the mean absolute errors of random effects. True edges are in bold.}
\label{tab:sim}
\end{table}

%% file: graph.tex
\begin{figure}
\vspace*{-15mm}
\centering\includegraphics[scale=1, angle=270]{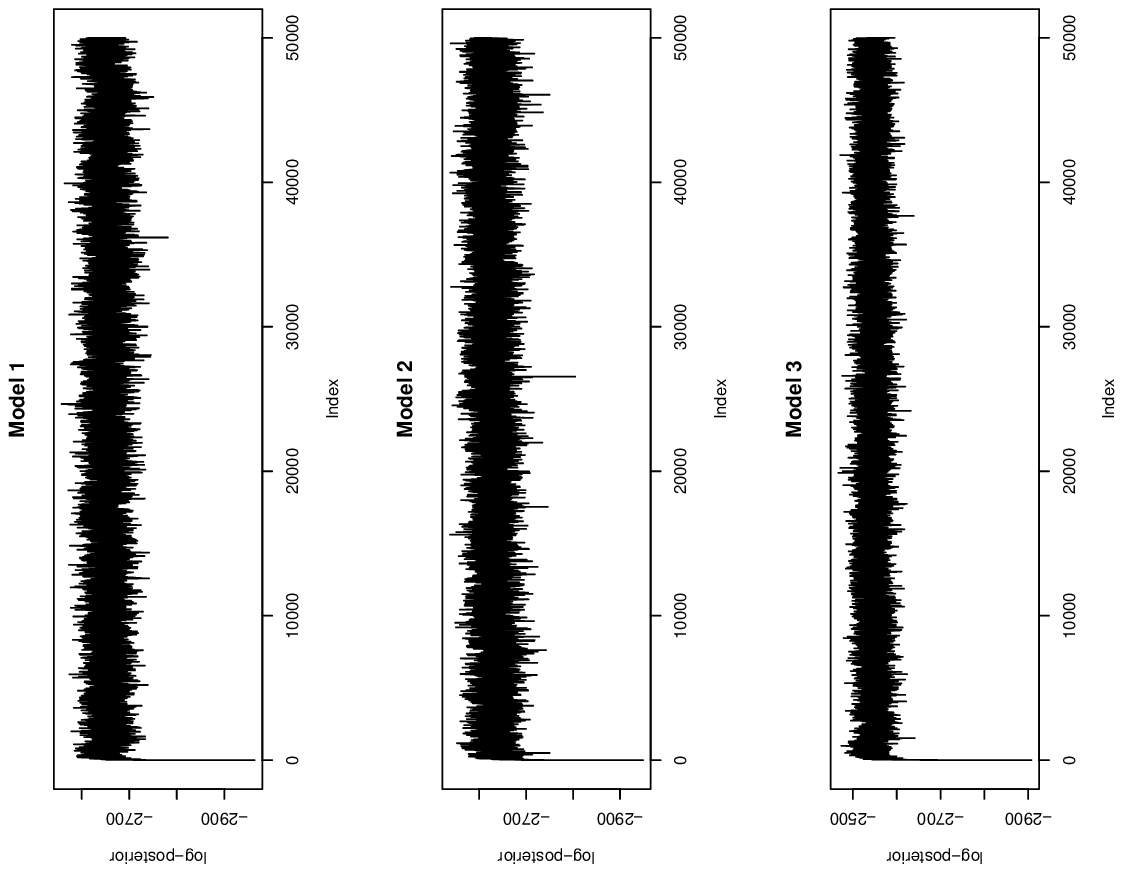}
\caption{Convergence of the log joint posterior under the three models (first 50,000 iterations).}
\label{fig:conv}
\end{figure}

\begin{figure}
\vspace*{-15mm}
\centering\includegraphics[scale=1]{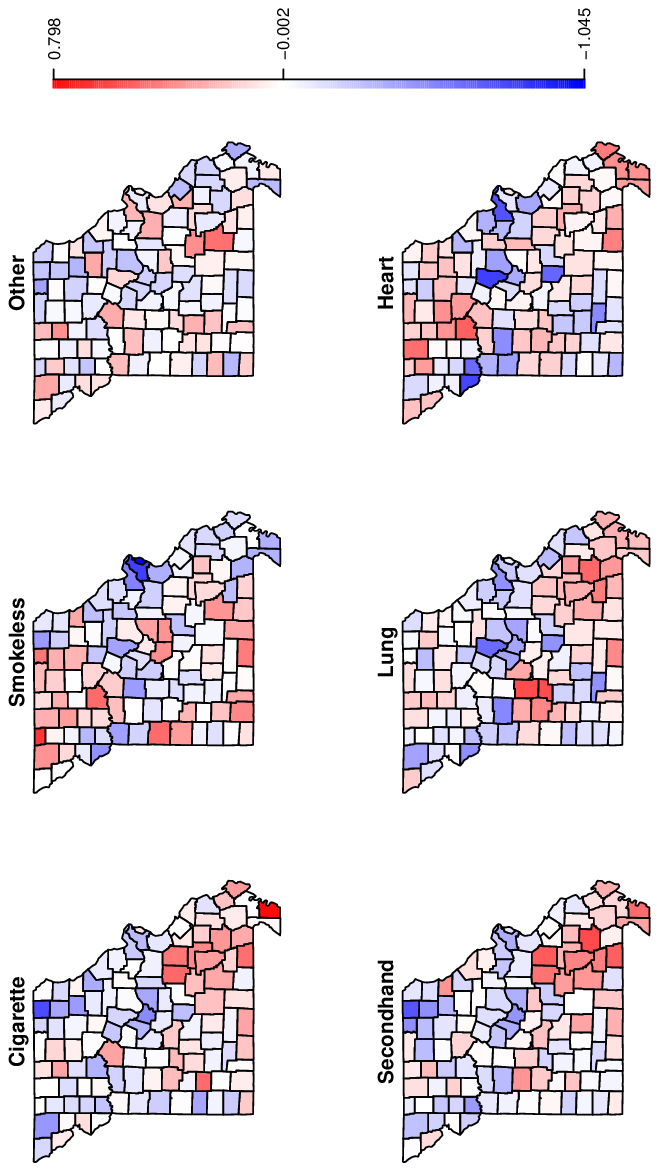}
\caption{Posterior means of spatial random effects $u_{ij}$ under Model 1.}
\label{fig:map}
\end{figure}

\begin{figure}
\vspace*{-15mm}
\centering\includegraphics[scale=1]{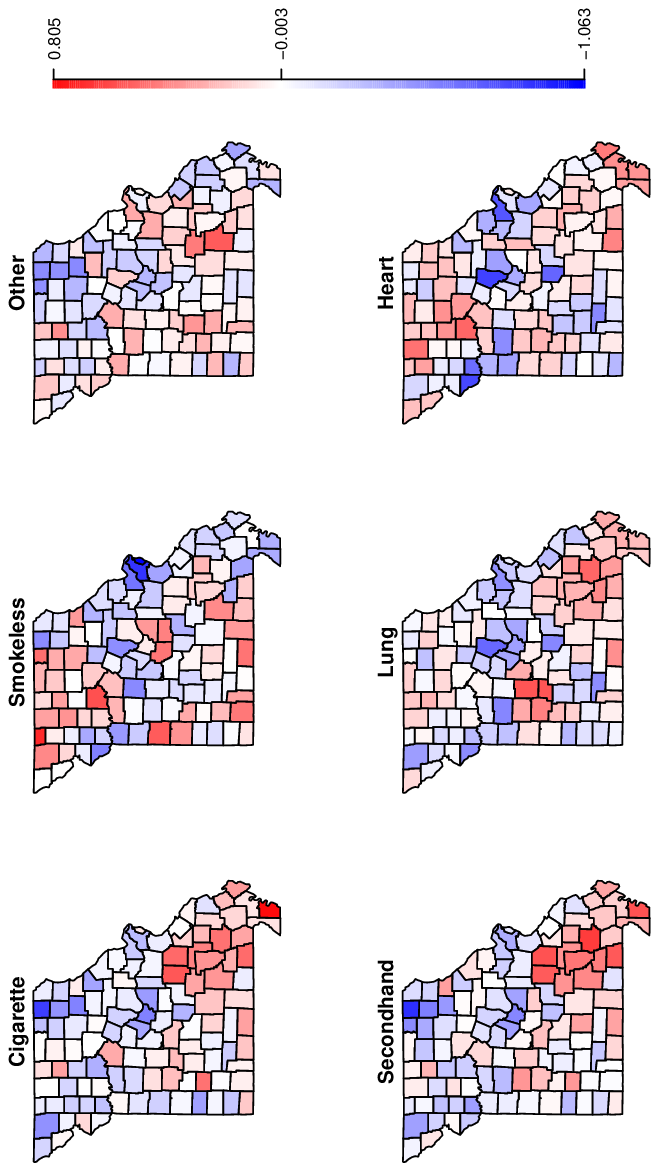}
\caption{Posterior means of spatial random effects $u_{ij}$ under Model 2.}
\label{fig:map}
\end{figure}

\begin{figure}
\vspace*{-15mm}
\centering\includegraphics[scale=1]{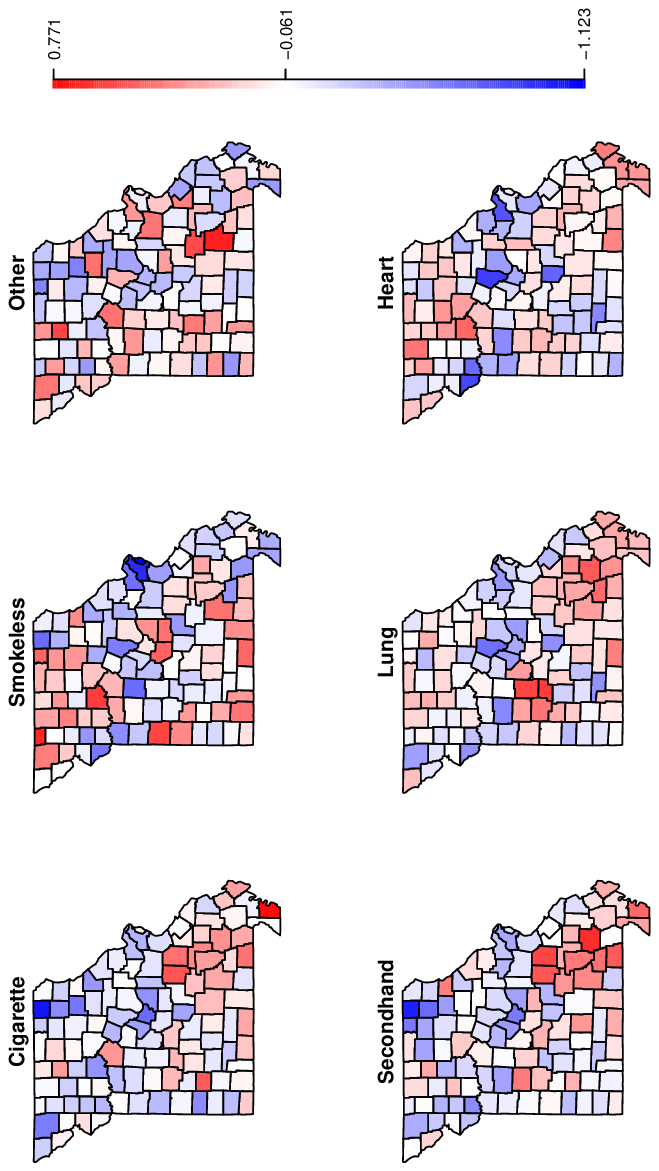}
\caption{Posterior means of spatial random effects $u_{ij}$ under Model 3.}
\label{fig:map}
\end{figure}